\documentclass[letterpaper, 10 pt, conference]{IEEEtran}  

\IEEEoverridecommandlockouts                              

\title{\LARGE \bf
Information Diffusion  and Preferential Attachment in a Network of Large Language Models
}

\usepackage{graphicx} 

\usepackage{amsmath,amsfonts,amssymb,amsthm}
\usepackage{bbm}
\usepackage{xcolor}
\usepackage{algorithm,algorithmic}
\usepackage{tikz}
\usepackage{dsfont}
\tikzset{every picture/.style={line width=0.75pt}} 

\newtheorem{theorem}{Theorem}
\newcommand{\numagents}{N}
\newcommand{\llma}{LLM}
\newcommand{\llmas}{{LLMs}}

\newcommand{\vertices}{\mathcal{V}}

\newcommand{\adjacency}{\mathbf{A}}
\newcommand{\nodeone}{i}
\newcommand{\nodetwo}{j}
\newcommand{\numrounds}{K}
\newcommand{\roundparticipant}{\mathcal{G}}
\newcommand{\timeindex}{k}
\newcommand{\observation}{y}
\newcommand{\observationspace}{\mathcal{Y}}

\newcommand{\state}{x}
\newcommand{\statespace}{\mathcal{X}}
\newcommand{\stateestimate}{\hat{\state}}
\newcommand{\latentstate}{z}
\newcommand{\indicator}{\mathds{1}}
\newcommand{\probability}{\mathbb{P}}
\newcommand{\neighbours}{\mathcal{N}}
\newcommand{\fractiontruthful}{\boldsymbol{\rho}}
\newcommand{\estimate}[1]{\hat{#1}}
\newcommand{\dynamicequation}{\mathbf{F}}

\newcommand{\timeindexalt}{t}
\newcommand{\truthful}{T}
\newcommand{\dontknow}{D}
\newcommand{\hallucinating}{H}

\newcommand{\latentspace}{\mathcal{Z}}
\newcommand{\probabilitylink}{\theta}
\newcommand{\expectation}{\mathbb{E}}
\newcommand{\avgprobabilitytransition}[2]{G_{#1#2}}
\newcommand{\degreedist}{q}
\newcommand{\degreeidx}{l}
\newcommand{\functiontransition}{\kappa}
\newcommand{\idx}{i}
\newcommand{\contextlength}{U}
\newcommand{\idxalt}{j}
\newcommand{\rategainconn}{\psi}
\newcommand{\rateloseconn}{\upsilon}
\newcommand{\degreedistvect}{\mathbf{\degreedist}}
\newcommand{\degreechangematrix}{\mathbf{H}}

\newcommand{\solutionalgebraic}{\Psi}
\newcommand{\transpose}{\mathsf{T}}
\newcommand{\control}{u}
\newcommand{\costfunction}{c}
\newcommand{\timehorizon}{L}
\newcommand{\constants}{\xi}
\newcommand{\orderof}{\mathsf{O}}
\newcommand{\erdosrenyi}{Erdős–Rényi }

\newcommand{\grade}{r}
\newcommand{\gradingfunction}{g}

\newcommand{\degreedistspace}{\mathcal{Q}}
\newcommand{\controlspace}{\mathcal{U}}
\newcommand{\gradingregime}{\mathcal{G}}
\newtheorem{proposition}{Proposition}

\newcommand{\gradevector}{\mathbf{\grade}}

\newcommand{\idxalttwo}{m}
\newcommand{\expectedgrade}{\mu}
\newcommand{\lyuponov}{V}

\author{Adit Jain, \textit{Student Member, IEEE}, Vikram Krishnamurthy, \textit{Fellow, IEEE},  Yiming Zhang  
\thanks{This work was supported by NSF under grant CCF-2312198.}
\thanks{A. Jain, V. Krishnamurthy, and Y. Zhang are with the School of Electrical and Computer Engineering, Cornell University, Ithaca, NY, USA (14853)
        {\tt\small aj457@cornell.edu}}%
}
\date{February 2025}

\begin{document}

\maketitle
\thispagestyle{empty}
\pagestyle{empty}

\begin{abstract}
This paper models information diffusion in a network of Large Language Models (LLMs) that is designed to answer queries from distributed datasets, where the LLMs can hallucinate the answer. We introduce a two-time-scale dynamical model for the centrally administered network, where opinions evolve faster while the network's degree distribution changes more slowly. Using a mean-field approximation, we establish conditions for a locally asymptotically stable equilibrium where all LLMs remain truthful. We provide approximation guarantees for the mean-field approximation and a singularly perturbed approximation of the two-time-scale system. To mitigate hallucination and improve the influence of truthful nodes, we propose a reputation-based preferential attachment mechanism that reconfigures the network based on LLMs' evaluations of their neighbors. Numerical experiments on an open-source LLM (LLaMA-3.1-8B) validate the efficacy of our preferential attachment mechanism and demonstrate the optimization of a cost function for the two-time-scale system.
\end{abstract}
\section{Introduction}
Networks of large language models (\llmas) have gained significant practical importance due to the increasing amount of online content generated and analyzed by LLMs~\cite{accessieee}. However, inaccuracies or absence of grounding in context can lead LLMs to produce hallucinations or irrelevant communication~\cite{sahoo2024comprehensive}. Motivated by these challenges, we consider a network comprising $\numagents$ LLMs tasked with estimating an underlying state from high-dimensional textual and visual observations through mutual interactions. Specifically, we address two fundamental questions:
\begin{enumerate}
\item How can the connectivity within an LLM network be adjusted to enhance the influence of truthful nodes?
\item How can we optimize the descriptiveness of LLM communications to minimize both unnecessary exchanges and hallucinations?
\end{enumerate}
To come up with an analytically tractable model, we first characterize the information diffusion dynamics using a mean-field approximation. We specifically study this problem in the context of \llmas, which can interact with to make inferences and extract relevant information. Further, we consider a two-time-scale system where the \llmas' latent state of the LLMs (if the LLM is truthful, hallucinating, or does not know), $\fractiontruthful$ evolves on a faster time scale, and the degree distribution of the network $\degreedistvect$ evolves on the faster time-scale. We provide an analytical singularly perturbed approximation of the above system, which can be helpful for future research. We propose a preferential attachment mechanism to readjust the adjacency matrix dynamically on a slower time scale. Finally, we propose optimizing for the control variable $\control$ so that an expectation of a cost function $\costfunction$ is minimized to increase communication efficiency and convergence rate.

\subsection{Motivation for Analyzing Network Of LLMs}
Large Language Models\footnote{We use LLMs to refer to vision language models as well.} are neural networks with $10^{9}-10^{12}$ parameters which are trained on a huge corpus of data ($\sim 10^{13}$ tokens) to process multimodal, especially textual information and generate textual output. They are already deployed in many practical applications, including code generation, customer service, and document analysis in healthcare, finance, and academia. However, in spite of a lot of active research, LLMs still have certain shortcomings; namely, they struggle to reliably perform long-context tasks and often hallucinate in the presence of ungrounded or misgrounded information. Additionally, they often are used in sensitive applications where maintaining privacy is important. We therefore motivate studying network dynamics of interacting LLMs with the following, 

\textit{Reliable Long-Context Understanding: }
Even trillion parameter LLMs demonstrate performance degradation as context (input) length increases beyond certain thresholds~\cite{gupta-etal-2024-systematic}, and this is due to the inherent nature of the self-attention mechanism and the training data. There are a few very long context window LLMs, but testing them exhaustively on their reliability is a challenging task~\cite{longcontexteval}. Therefore, we propose and analyze a network of LLMs, each of which can reliably operate in a different and smaller context and then communicate to perform information processing. 

\textit{Privacy: }  Centralized LLMs require complete data access, violating privacy constraints in sensitive domains like healthcare and finance~\cite{song-etal-2024-securesql}. A network of LLMs enables information processing without centralizing raw data. LLMs exchange only derived insights rather than original documents, preserving the sovereignty of the data. Further, the type of data shared between LLMs can be defined. 

\textit{Enhanced Robustness: }
Individual LLMs frequently hallucinate, i.e., generate convincing but factually incorrect information, particularly beyond their training data or the information available on the internet~\cite{sahoo2024comprehensive}. Using reputation scores and trust-based reconfiguration, our proposed dynamic network readjustment improves factual reliability without requiring model retraining. Further, by incorporating LLMs trained on specialized datasets, we can have a heterogeneous network that can handle a diverse set of tasks. Finally, we offer a theoretical framework for controlled information propagation, which is absent in the current literature of interacting LLMs. Additionally, the computational complexity for inference in the context of length $\contextlength$ is $\orderof(\contextlength^2)$. Therefore, it is computationally more efficient to chunk the content into $\numagents$ chunks and process it reliably using a model with less ($50\times$) number of parameters~\cite{subramanian2025smalllanguagemodelsslms}.

\subsection{Related Work}
There has been significant recent work on practical orchestration of multi-LLM coordination across various domains, including question answering systems and processing of long-context documents using LLM chains~\cite{seabra2024dynamicmultiagentorchestrationretrieval,autotqa}. However, the current literature lacks theoretical rigor and techniques for dynamically readjusting LLM networks, particularly at scale—a gap this paper addresses. Network science has extensively studied mechanisms for controlling information flow~\cite{López-Pintado2016,s16040481,li2017mechanismdesignsocialnetworks}, which we extend by investigating
3 latent states. Researchers have examined consensus formation in two-time-scale systems~\cite{consensusformationtwotimescale} and sparse input controllability within \erdosrenyi networks~\cite{joseph2021controllability}, but these studies focused primarily on human networks or physical sensor systems. LLM networks demand a fresh perspective given their unique combination of cognitive abilities and steerability~\cite{bhargava2024whatsmagicwordcontrol}. Additionally, in a centrally controlled setting, one can also tweak the adjacency matrix of the network of LLMs. In similar spirit~\cite{accessieee} looked at controlling Bayesian social learning in LLMs, where the results were restricted to a sequential network of LLMs.

 

\subsection{Main Contributions}
\begin{enumerate}
    \item We formulate the problem of distributed state estimation in a network of large language models and analyze the mean-field field approximation of the information propagation dynamics. In Theorem~\ref{thm:stability}, we provide sufficient conditions for the state with all truthful LLMs to be in a locally asymptotically stable equilibrium. 
        \item We analyze the complete system as a two-time-scale system, where the information diffusion is on the faster time scale, and the network reconfiguration is on the slower time scale. We reduce the two-time-scale system to a singularly perturbed system whose approximation error is characterized in Theorem~\ref{thm:singular_perturbation}. 
    \item Using the analytic capabilities of the \llmas, we propose a reputation score computed using grades provided by an LLMs' neighbors. We propose a preferential attachment mechanism in Algorithm~\ref{alg:trustrankbasedreadjustment} to adaptively reconfigure the network using the reputation score computed after phase one. In Proposition~\ref{thm: concentration on grading} we show that our protocol ensures truthful LLMs have more outgoing edges under reasonable conditions on the computed reputation scores.  
    \item We show the efficacy of our proposed method on Llama 3.1 8B numerical experiments on synthetically generated datasets for improved distributed inference to reduce hallucination. We also show how a cost function for a network of LLMs can be minimized. 
\end{enumerate}
\textbf{Notation:} We denote vectors by bold-face small case letters, $\mathbf{x}$, matrices by bold-face capital case letters, $\mathbf{A}$ and $\mathbf{x}^\transpose$ denotes the transpose. $\mathbf{x}_j$ denotes the $j$-th element of the vector. $\probability$ and $\expectation$ denote probability and expectation with respect to the appropriate probability measure. $[n]$ denotes the sequence $1,\dots,n$.
\section{Network of LLMs for Distributed Inference}
This section develops mathematical abstraction for LLMs communicating over a dynamic network. We first develop a mean-field approximation for the dynamics of information diffusion in LLMs performing state estimation.  We provide a singularly perturbed approximation for the two-time scale system, which can be used to analyze the behavior of the combined system tractably.  Further, we propose an algorithm that a central controller can implement to dynamically adjust the network by evaluating the LLMs using their neighbors.
\subsection{System Model}\label{sec:sectionmodel}

\begin{figure}
    \centering
    \vspace{2mm}
    \resizebox{\columnwidth}{!}{
\begin{tikzpicture}[x=0.75pt,y=0.75pt,yscale=-1,xscale=1]

\draw   (276,250) -- (458,250) -- (458,350) -- (276,350) -- cycle ;
\draw   (296,270) -- (426,270) -- (426,330) -- (296,330) -- cycle ;
\draw    (240,300) -- (278,300) ;
\draw [shift={(280,300)}, rotate = 180] [color={rgb, 255:red, 0; green, 0; blue, 0 }  ][line width=0.75]    (10.93,-3.29) .. controls (6.95,-1.4) and (3.31,-0.3) .. (0,0) .. controls (3.31,0.3) and (6.95,1.4) .. (10.93,3.29)   ;
\draw    (426,300) -- (488,300) ;
\draw [shift={(490,300)}, rotate = 180] [color={rgb, 255:red, 0; green, 0; blue, 0 }  ][line width=0.75]    (10.93,-3.29) .. controls (6.95,-1.4) and (3.31,-0.3) .. (0,0) .. controls (3.31,0.3) and (6.95,1.4) .. (10.93,3.29)   ;
\draw    (240,340) -- (274,340) ;
\draw [shift={(276,340)}, rotate = 180] [color={rgb, 255:red, 0; green, 0; blue, 0 }  ][line width=0.75]    (10.93,-3.29) .. controls (6.95,-1.4) and (3.31,-0.3) .. (0,0) .. controls (3.31,0.3) and (6.95,1.4) .. (10.93,3.29)   ;
\draw    (240,260) -- (274,260.94) ;
\draw [shift={(276,261)}, rotate = 181.59] [color={rgb, 255:red, 0; green, 0; blue, 0 }  ][line width=0.75]    (10.93,-3.29) .. controls (6.95,-1.4) and (3.31,-0.3) .. (0,0) .. controls (3.31,0.3) and (6.95,1.4) .. (10.93,3.29)   ;
\draw  [dash pattern={on 0.84pt off 2.51pt}] (30,230) -- (250,230) -- (250,370) -- (30,370) -- cycle ;

\draw (221,292.4) node [anchor=north west][inner sep=0.75pt]    {$y_{k}$};
\draw (191,293.4) node [anchor=north west][inner sep=0.75pt]    {$x_{k}$};
\draw (201,294.4) node [anchor=north west][inner sep=0.75pt]    {$\sim $};
\draw (308,282) node [anchor=north west][inner sep=0.75pt]   [align=left] {\begin{minipage}[lt]{77.03pt}\setlength\topsep{0pt}
Large Language
\begin{center}
Model
\end{center}

\end{minipage}};
\draw (221,252.4) node [anchor=north west][inner sep=0.75pt]    {$u$};
\draw (41,262) node [anchor=north west][inner sep=0.75pt]  [font=\footnotesize] [align=left] {Control \ (System Prompt)};
\draw (41,289) node [anchor=north west][inner sep=0.75pt]  [font=\footnotesize] [align=left] {Private Observation\\(User Prompt)};
\draw (41,328) node [anchor=north west][inner sep=0.75pt]  [font=\footnotesize] [align=left] {Previous Action\\of Neighbours (Context)};
\draw (36,237) node [anchor=north west][inner sep=0.75pt]   [align=left] {LLM Input };
\draw (431,228) node [anchor=north west][inner sep=0.75pt]   [align=left] {LLM Output};
\draw (361,330.4) node [anchor=north west][inner sep=0.75pt]    {$z_{k}$};
\draw (501,290.4) node [anchor=north west][inner sep=0.75pt]    {$\hat{x}_{k}^{i}$};
\draw (181,330.4) node [anchor=north west][inner sep=0.75pt]    {$\hat{x}_{k-1}^{i} ,\hat{x}_{k-1}^{-i}$};

\end{tikzpicture}
}
        \vspace{-10mm}
    \caption{System Model of a Single LLM: At each time step $\timeindex$, each LLM receives a composite input consisting of a private observation from the true state $\state$, answers from the previous round of the LLM and its neighbors, and the system prompt (which acts as the control). The LLM outputs either an estimate of the state or ``does not know''. If the private information is non-informative, the LLM can either say it does not know or can make up a state estimate. }
        \vspace{-7mm}

    \label{fig:systemmodel}
\end{figure}
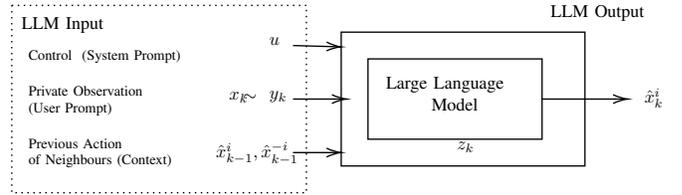
We consider a network of LLMs, each of which acts like a dynamic sensor, and provide estimates of a state from a given text observation as illustrated in Figure~\ref{fig:systemmodel}.
There are $\numagents$ \llmas\ which communicate with each other using a network where the network adjacency matrix is denoted by $\adjacency \in \{0,1\}^{\numagents\times\numagents}$. In our setup, we consider asymmetric interactions where if $\adjacency_{\nodeone,\nodetwo} = 1$, there is an edge from $i$ to $j$ and the \llma\  $\nodeone$ is influenced by \llma\ $\nodetwo$, i.e., it considers the previous observations of \llma\ $\nodetwo$ while processing an output (explained next).  We denote the influencing neighbors of \llma\ $\nodeone$ by $\neighbours(\nodeone) = \{\nodetwo| \adjacency_{\nodeone,\nodetwo} = 1 \}$. The \llmas\ interact over $\numrounds$ rounds. At each time $\timeindex$, let denote $\roundparticipant_\timeindex\subseteq [\numagents]$ the set of \llmas\ which interact at time $\timeindex \in 1,2,\dots$. Each \llma\ $\nodeone$ has a private observation $\observation^\nodeone \in \observationspace$, where $\observationspace$ is a high-dimensional observation space like the space of all text or images. Our framework does not assume a certain network structure, e.g., sampled as a \erdosrenyi\ or from a power-law degree distribution, offering flexibility to the LLM network designer.

The collective aim of the \llmas\ is to communicate and estimate an underlying unknown state $\state\in\statespace$ (e.g., the correct answer to a question), where $\statespace$ is the state space. Each \llma\ also has an unknown latent state $\latentstate^\nodeone_\timeindex \in \latentspace = \{ -1 = \texttt{does not know} \ (\dontknow),0 = \texttt{hallucinating} \ (\hallucinating),1 = \texttt{correct} \ (\truthful)\}$. At time $\timeindex$, let $\observation^\idx_\timeindex \in \observationspace$ denote an observation sampled (e.g., LLM's answer along with explanation) from an LLM $\idx$ in latent state $\latentstate^\nodeone_\timeindex$ as $\observation^\idx_\timeindex\sim \idx|\latentstate^\nodeone_\timeindex$, where $\observationspace$ is a high-dimensional text space.  At each round, the LLM is provided with the previous observations of its neighbors $\{\observation^\nodetwo_{\timeindex-1}:  \nodetwo \in \neighbours(\nodeone) \}$ and their previous observation $\observation^\nodeone_\timeindex$.  Each LLM $\nodeone$ is polled at the end of round $\timeindex$ for their estimate of the state, $\stateestimate^\nodeone_\timeindex \in \statespace \cup \{-1=\texttt{`do not know'}\}$ (e.g., LLM's answer).  The latent state is defined as $\latentstate^\nodeone_\timeindex(\stateestimate^\nodeone_\timeindex,\state) = \indicator(\stateestimate^\nodeone_\timeindex=\state) - \indicator(\stateestimate^\nodeone_\timeindex=-1)$. 
Further, the network controller can also control the behavior of the \llmas\ through a control $\control \in \controlspace$, where $\controlspace$ is a finite set of controls. In practice, this would be limited to how the LLMs generate content, specifically how much computing is available for reasoning and communicating their answer.

\subsection{Mean Field Approximation of Network Dynamics}
We model the spread of information in the network of \llma, as a Markov process, where the latent state of each LLM depends on its previous state, the previous state of its neighbors, and the private observation. 
Note that we analyze the spread of information in the latent space because we treat the true state as unknown but fixed.  Therefore, the conditional probabilities are the same since, given the true state, the latent variable is a deterministic map from the estimated state to the latent state.
We are interested in truthful latent state $1$ and specifically the quantity,
    $\estimate{\fractiontruthful_\truthful} =  \frac{1}{\numagents}\sum_1^\numagents \indicator(\latentstate^\nodeone=\truthful)$. 
We study the case when number of agents and to get a tractable analysis and present theoretical results, look at the quantity $\fractiontruthful_\latentstate$, which is the population average of latent state $\latentstate$. 

Let $\fractiontruthful_{\timeindex,\latentstate}^\degreeidx$ denote the relative density of latent state $\latentstate$ in nodes with in-degree $\degreeidx$ at time $\timeindex$, therefore $\sum_{\latentstate\in\latentspace} \fractiontruthful_{\timeindex,\latentstate}^\degreeidx  = 1$ for all times $\timeindex$ and out-degrees $\degreeidx \in [\numagents]$. Let $\fractiontruthful^\degreeidx_0$ be the initial latent distribution for LLMs with in-degree $\degreeidx$.  Let $\degreedistvect$ denote the in-degree distribution 
from the space of degree distributions $\degreedistspace$ . 
Let $\probabilitylink_\latentstate$ be the probability of a randomly sampled link that starts is directed to a node with latent state $\latentstate$. In general asymmetric networks it is difficult to obtain deriving the expression for $\probabilitylink_\latentstate$, it depends on the the out-degree distribution $\degreedistvect$ and on the latent state distribution with nodes with different out-degrees~\cite{lopez-pintado_diffusion_2008}. However note that if $\fractiontruthful=[1,0,0]^\transpose$, then $\probabilitylink_\truthful = 1$ and $\probabilitylink_\hallucinating = 0$.

$\avgprobabilitytransition{\latentstate_1}{\latentstate_2}^\degreeidx(\degreedistvect,\control)$ is the average transition probability given by the following expression,
\begin{align}\label{eq:G}
\begin{split}
    \avgprobabilitytransition{\latentstate_1}{\latentstate_2}^\degreeidx&(\degreedistvect,\control)= \sum_{i=0}^{\degreeidx}\sum_{\idxalt=0}^{\degreeidx-i} \functiontransition_{\latentstate_1,\latentstate_2}(\control,\degreeidx,\idx,\idxalt) \binom{\degreeidx}{\idx,\idxalt} \\&(\probabilitylink_\truthful(\degreedistvect))^\idx (\probabilitylink_\hallucinating(\degreedistvect))^\idxalt (1 - \probabilitylink_\truthful(\degreedistvect)-\probabilitylink_\hallucinating(\degreedistvect))^{\degreeidx-\idx -\idxalt},
    \end{split}
\end{align}
where $\functiontransition_{\latentstate_1,\latentstate_2}(\control,\degreeidx,\idx,\idxalt)$ denotes the probability of a LLM with degree $\degreeidx$ transition from state $\latentstate_1$ to state $\latentstate_2$ given that it has $\idx$ truthful and $\idxalt$ hallucinating neighbours.

Therefore, we can derive the evolution of the latent state distribution for a fixed degree distribution $\degreedistvect$ and control $\control$, 
\begin{align}\label{eq:degreebaseddynamiceqn}
    \frac{d}{d\timeindexalt}\fractiontruthful^\degreeidx(\timeindexalt)  = \dynamicequation^\degreeidx(\degreedistvect,\control) \fractiontruthful^\degreeidx(\timeindexalt),
\end{align}
where $\dynamicequation^\degreeidx$ is given by,
\begin{align*}
\dynamicequation^\degreeidx_{\latentstate_1,\latentstate_2}(\degreedistvect,\control) &= \begin{cases}
\avgprobabilitytransition{\latentstate_1}{\latentstate_2}^\degreeidx(\degreedistvect,\control), \ \latentstate_1 \neq \latentstate_2   \\
-\sum_{\latentstate_2^\prime\in\latentspace, \latentstate_2^\prime\neq\latentstate_1}\avgprobabilitytransition{\latentstate_1}{\latentstate_2^\prime}^\degreeidx(\degreedistvect,\control), \latentstate_1 = \latentstate_2
      \end{cases}. 
\end{align*}

\begin{enumerate}
    \item[(A1)] \(\kappa_{T H}(u, l, l, 0) = 0\) and \(\kappa_{T D}(u, l, l, 0) = 0\) for all \(l\),
    \item[(A2)] \(\kappa_{H T}(u, l, i, j) > 0\) and \(\kappa_{D T}(u, l, i, j) > 0\) for all \(i, j\) with \(i \geq l - \delta l\), where \(\delta l\) is a small positive integer.
    \item[(A3)] There exists small $\epsilon$ such that if  $\fractiontruthful_\hallucinating^\degreeidx,\fractiontruthful_\dontknow^\degreeidx<\frac{\epsilon}{2} \ \forall \degreeidx$, then for all $\degreedistvect\in\degreedistspace$, $\probabilitylink_{\truthful}>1-\epsilon$ and $\probabilitylink_\hallucinating<\frac{\epsilon}{2}$.
\end{enumerate}
Assumption A1 ensures that when all the neighbors are truthful, then the LLMs will not change their latent state from truthful to hateful or do not know. Assumption A2 ensures that if there are enough truthful neighbors, then the probability of transitioning to truthful is non-zero. 
Next, we show our first result, which shows that under these assumptions, a locally asymptotically stable equilibrium exists with all truthful LLMs. Assumption A3 ensures that if the proportion of hallucinating or dont-know nodes of any out-degree is very small, then the probability of sampling an edge to a halluncinating and dont-know node will also be very small. 
\begin{theorem}\label{thm:stability}
Consider a network of large language models (LLMs) with mean-field dynamics described by the system~\eqref{eq:degreebaseddynamiceqn}
where \(\rho^l = [\rho_T^l, \rho_H^l, \rho_D^l]^T\) denotes the state distribution for LLMs of degree \(l\), with \(\rho_T^l + \rho_H^l + \rho_D^l = 1\), \(\mathbf{q}\) is the degree distribution of the network, and \(u \in \mathcal{U}\) is a control parameter. The equilibrium state \(\rho^l_* = [1, 0, 0]^T\) for all degrees \(l\) (i.e., \(\rho_T^l = 1\), \(\rho_H^l = 0\), \(\rho_D^l = 0\)) is locally asymptotically stable if there exists a control \(u \in \mathcal{U}\) satisfying A1, A2 and A3. 
\end{theorem}
\textit{Proof in Appendix. }

The above theorem ensures the existence of a locally stable equilibrium, given that the control $\control$ can control the propagation. In the case of LLMs, where the control is either controlling how much they are allowed to communicate or how comprehensive their instruction prompt is, the above theorem guarantees the existence of a stable equilibrium at $\fractiontruthful=(1,0,0)^\transpose$ if the control can control the behavior of an individual LLM. However, note that although such a control may exist, it may be too costly to implement. Therefore, this paper aims to optimize a cost function so that a reasonable consensus can be reached efficiently. 

In contrast to the result from~\cite{lopez-pintado_diffusion_2008}, the above result only provides sufficient conditions, this is primarily because deriving necessary conditions for a $3$ state system is significantly more challenging. Further condition A3 is satisfied automatically if the network is symmetric.

We now present the following well-known result, the proof of which can be found in~\cite{krishnamurthy_partially_2016}, which shows how good a mean-field approximation is to the true network dynamics. 
\begin{theorem}[\cite{krishnamurthy_partially_2016}](\textit{Approximation Result using Concentration inequality}) Let $\fractiontruthful_\latentstate^\degreeidx(\timeindex)$ denote the proportion of LLMs with degree $\degreeidx$ in state $\latentstate$ at time $\timeindex$ which evolves as~\eqref{eq:degreebaseddynamiceqn} with the initial state $\fractiontruthful_\latentstate^\degreeidx(0)$. Further let $\hat{\fractiontruthful}_\latentstate^\degreeidx(\timeindex)$ denote the empirical proportion in a network with $\numagents$ LLMs. Then for any $\epsilon>0$ and time horizon $\timehorizon= \orderof(\numagents)$, there exists constants $\constants_1,\constants_2>0$ such that,
\begin{align}
    \probability\left(\max_{0\leq \timeindex \leq \timehorizon } \Vert \fractiontruthful_\latentstate^\degreeidx(\timeindex) - \hat{\fractiontruthful}_\latentstate^\degreeidx(\timeindex) \Vert \geq \epsilon\right) \leq \constants_1 \exp(-\constants_2\epsilon^2\timehorizon).
\end{align}
\end{theorem}
Theorem 2 implies that for a large enough number of LLMs, the mean-field approximation can be analyzed instead of the true dynamics for a reasonably long horizon. This is especially useful when the network dynamics evolve on a time scale, which is much faster, as is the case in the two-time scale system we discuss, where the slower time-scale readjusts the network's degree distribution.

The network dynamics are run on a faster time scale, and the network is reconfigured on a slower time scale. 
Let the dynamics of the degree distribution on the slower time-scale, 
\begin{align}\label{eq:degreedynamicequation}
    \frac{d\degreedistvect(\timeindexalt)}{d\timeindexalt} = \degreechangematrix(\fractiontruthful)\degreedistvect(\timeindexalt),
\end{align}
Therefore 
we have the following two-time-scale system which combines the two dynamical systems of~\eqref{eq:degreebaseddynamiceqn} and~\eqref{eq:degreedynamicequation},
\begin{align}\label{eq:two-timescalesystem}
    \begin{split}
        \varepsilon\frac{d\fractiontruthful}{d\timeindexalt} &= \dynamicequation(\degreedistvect) \fractiontruthful \\
         \frac{d\degreedistvect}{d\timeindexalt} &= \degreechangematrix(\fractiontruthful)\degreedistvect,
    \end{split}
\end{align}
where $\varepsilon$ is the scaling factor across the time-scales. 

As $\varepsilon\to 0$ the faster time-scale system can be approximated by a algebraic equation, $\dynamicequation(\degreedistvect) \fractiontruthful = 0$. 
Let $\fractiontruthful = \solutionalgebraic(\degreedistvect)$ be the solution of this algebraic equation along with the constraint on $\fractiontruthful$. Plugging the solution into the slower time-scale system equation of~\eqref{eq:two-timescalesystem}, we obtain the following singularly perturbed time-scale system~\cite{khalil2002nonlinear},
\begin{align}\label{eq:singlyperturbedtimescalesystem}
    \frac{d\degreedistvect}{d\timeindexalt} &= \degreechangematrix(\solutionalgebraic(\degreedistvect))\degreedistvect.
\end{align}
This is a useful abstraction and can be used to analyze the original system, where the approximation error can be bounded under regularity conditions on the dynamic equations and convergence to stable equilibria exponentially fast. This can be formalized in the following result, which is a restated version of~\cite[Theorem 8.1]{khalil2002nonlinear}.  
\begin{theorem}[\cite{khalil2002nonlinear}]
\label{thm:singular_perturbation}
(Approximation of two-time-scale system as a  singularly perturbed system) Let the distributed inference dynamics in a network of \llmas\ be governed by the two-time-scale system~\eqref{eq:two-timescalesystem}.
(\textit{A1}) Let the real parts of all eigenvalues of the Jacobian $\frac{d\dynamicequation(\degreedistvect)\fractiontruthful}{d\timeindexalt}\mid_{\solutionalgebraic(\degreedistvect)}$ are negative for all $\degreedistvect \in \degreedistspace$. (\textit{A2}) Further, assume that the functions $\rategainconn$ and $\rateloseconn$ are Lipschitz continuous and bounded in $\fractiontruthful$. 
Let $\degreedistvect^*(\timeindexalt)$ denote the solution to the reduced system~\eqref{eq:singlyperturbedtimescalesystem} and $\fractiontruthful^*(\timeindexalt) = \solutionalgebraic(\degreedistvect^*(\timeindexalt))$ the corresponding quasi-steady state. 
Then for any $\timeindexalt \in [\timeindexalt_0, \mathsf{T}]$ and initial conditions satisfying $\|\fractiontruthful(0) - \solutionalgebraic(\degreedistvect(0))\|= \orderof(\varepsilon)$:
\begin{align*}
    \|\degreedistvect(\timeindexalt) - \degreedistvect^*(\timeindexalt)\| &\leq \constants_1\varepsilon \\
    \|\fractiontruthful(\timeindexalt) - \fractiontruthful^*(\timeindexalt)\| &\leq \constants_2\varepsilon
\end{align*}
for appropriate constants $\constants_1, \constants_2 > 0$ independent of $\varepsilon$.
\end{theorem} 
Theorem~\ref{thm:singular_perturbation}  simplifies the complex two-time-scale system into a more analytically tractable singular perturbation approximation with error bounds. 
Unlike traditional networks, LLMs have unique properties (hallucination tendencies, linguistic reasoning capabilities) that create distinct dynamics when connected. This framework provides analytical tools to understand these emergent behaviors. Future research can leverage this approximation to develop provable bounds on hallucination propagation and design network topologies that minimize false information spread while maximizing resource efficiency.

\subsection{Preferential Attachment for improved connectivity}
Since this paper deals with a centrally controlled network of LLMs, we propose periodically altering the network to ensure that more reliable nodes are connected. The theoretical results of the paper are useful for analysis but are often difficult to use in practice. To complement our theory, we propose a preferential attachment protocol that improves the network of LLMs on a slower time-scale. Since the LLMs are cognitive sensors, we use the ability to evaluate textual output to grade their neighboring LLMs and use these grades to reconfigure the network. In our numerical results, we demonstrate how such an evaluation-based preferential attachment is more useful than preferential attachment mechanisms based on different centrality measures computed solely based on network structure.  
\begin{algorithm}
    \begin{algorithmic}[1]
        \STATE  \textit{Input: } Adjacency Matrix $\adjacency$, Expected Grades $\expectedgrade_\truthful,\expectedgrade_\hallucinating,\expectedgrade_\dontknow$, Observations $\{\observation_\idx\}_{\idx=1}^\numagents$
        \STATE \textit{Output: } Updated Adjacency Matrix $\adjacency^\prime$
        
        \STATE \textit{Initalize: } $\gradevector_\idx = \gradingfunction_\idx(\observation_\idx) \ \forall \idx \in [\numagents]$, $\adjacency^\prime = \adjacency$ 
        \FOR{$\idxalttwo \in [\log(\numagents)]$ } 
        \IF{$ \{(\idx,\idxalt) | \gradevector_\idxalt >\frac{\expectedgrade_\truthful+\expectedgrade_\dontknow}{2}, \adjacency_{\idx\idxalt} = 0, |\neighbours(\idxalt)|\geq\frac{4\log(2\numagents)}{(\expectedgrade_\truthful-\expectedgrade_\dontknow)^2}  \} \neq \emptyset$}
        \STATE Sample $(\idx^\prime,\idxalt^\prime) \sim \{(\idx,\idxalt) | \gradevector_\idxalt >\frac{\expectedgrade_\truthful+\expectedgrade_\dontknow}{2}, \adjacency_{\idx\idxalt} = 0, |\neighbours(\idxalt)|\geq\frac{4\log(2\numagents)}{(\expectedgrade_\truthful-\expectedgrade_\dontknow)^2}  \}$ and set $\adjacency^\prime_{\idx^\prime,\idxalt^\prime} = 1$
        \ENDIF
        \IF{$ \{(\idx,\idxalt) | \gradevector_\idxalt <\frac{\expectedgrade_\dontknow-\expectedgrade_\hallucinating}{2}, \adjacency_{\idx\idxalt} = 0, |\neighbours(\idxalt)|\geq\frac{4\log(2\numagents)}{(\expectedgrade_\dontknow-\expectedgrade_\hallucinating)^2}  \} \neq \emptyset$}
        \STATE Sample $(\idx^\prime,\idxalt^\prime) \sim \{(\idx,\idxalt) | \gradevector_\idxalt <\frac{\expectedgrade_\hallucinating+\expectedgrade_\dontknow}{2}, \adjacency_{\idx\idxalt} = 1, |\neighbours(\idxalt)|\geq\frac{4\log(2\numagents)}{(\expectedgrade_\dontknow-\expectedgrade_\hallucinating)^2}  \}$ and set $\adjacency^\prime_{\idx^\prime,\idxalt^\prime} = 0$
        \ENDIF

        \ENDFOR{}
    \end{algorithmic}
    \caption{Preferential Attachment based Readjustment}
    \label{alg:trustrankbasedreadjustment}
\end{algorithm}

Algorithm~\ref{alg:trustrankbasedreadjustment} implements a network readjustment mechanism using a preferential attachment protocol to control information propagation in LLM networks. It assigns reputation scores or grade $\grade_\idx$ for each node $\idx$ based on latent states $\latentstate_\idx$ using the function $\grade_\idx$. 
Then, at each of the $\log\numagents$ iterations, we sample one edge to add and one edge to delete. The sets that the sampling is done on in Step 6 and Step 9 consist of nodes with $\Omega(\log\numagents)$ neighbors to ensure that the grades assigned to these LLMs nodes have confidence at least $1-\frac{1}{\numagents}$. Steps 5 and 8 ensure that sampling is done from a non-empty set. We sample with replacement to show the theoretical results, but one could develop a sample without replacement version too. This approach modifies network connectivity based on trust metrics. We next describe the grading function $\grade_\idx$. This preferential attachment protocol with an initial power law distribution ensures the network remains sparse~\cite{Krishnamurthy2019}.
Our main ingredient in the network readjustment is a peer-driven evaluation mechanism. 

We define a grading aggregation function which takes text observation of $\observation_\idx$ and gives back an evaluation between interval $[0,1]$, $\gradingfunction_\idx(\observation_\idx) = \frac{1}{\neighbours_\idx}\sum_{\idxalt\in\neighbours_\idx}\gradingfunction(\observation_\idx)$, where  $\gradingfunction:\observationspace\to[0,1]$ is the individual grading function. 
The grading interval is partitioned into three disjoint intervals $\{\gradingregime_\truthful,\gradingregime_\hallucinating,\gradingregime_\dontknow\}$ each of which corresponds to one of the latent states. Let $\vertices_\latentstate = \{\idx | \latentstate_\idx = \latentstate \}, \latentstate\in\latentspace$ denote the set of LLMs which are truthful. We make the following assumptions: 

\begin{enumerate}
    \item[(B1)] The grading function $\gradingfunction$ gives noisy evaluation where the noise is independent of each other and is bounded by $1$. The expected grade $\expectedgrade_\latentstate = \expectation[\grade_\idx|\latentstate_\idx=\latentstate]$ for latent state $\latentstate$ is such that $\expectedgrade_\truthful > \expectedgrade_\dontknow > \expectedgrade_\hallucinating$. 
    \item[(B2)] The set $\{(\idx,\idxalt) | \gradevector_\idxalt >\frac{\expectedgrade_\truthful+\expectedgrade_\dontknow}{2}, \adjacency_{\idx\idxalt} = 0, |\neighbours(\idxalt)|\geq\frac{4\log(2\numagents)}{(\expectedgrade_\truthful-\expectedgrade_\dontknow)^2}  \}$ and the set $\{(\idx,\idxalt) | \gradevector_\idxalt <\frac{\expectedgrade_\dontknow+\expectedgrade_\hallucinating}{2}, \adjacency_{\idx\idxalt} = 0, |\neighbours(\idxalt)|\geq\frac{4\log(2\numagents)}{(\expectedgrade_\dontknow-\expectedgrade_\hallucinating)^2}  \}$ are non-empty.
\end{enumerate}
Assumption B1 ensures that the grades are informative of the states. Assumption B2 ensures that the sets of candidate LLM nodes being removed and added (whose evaluations are right with probability $1-\frac{1}{N}$) are non-empty.
\newcommand{\problarge}{p}
\begin{proposition}\label{thm: concentration on grading}
Let $\adjacency$ denote the adjacency matrix of large language model LLMs with the system model described in Section~\ref{sec:sectionmodel}. And let $\adjacency^\prime$ denote the readjusted adjacency matrix after running Algorithm~\ref{alg:trustrankbasedreadjustment}. Let Assumptions B1 and B2 hold and let $\problarge_\latentstate = \sum_{\degreeidx\geq \frac{4\log(2\numagents)}{|\expectedgrade_\latentstate-\expectedgrade_\dontknow|^2}}^{\numagents-1}\degreedistvect_\degreeidx \varrho^\degreeidx_\latentstate$, where $\degreedistvect$ is the in-degree distribution and $\varrho^\degreeidx_\latentstate$ is the latent state distribution for latent $\latentstate$ and LLMs with in-degree $\degreeidx$. Then after one run of readjustment using Algorithm~\ref{alg:trustrankbasedreadjustment}, the following bounds hold,
    \begin{align*}
\probability&\left(\sum_{\idx \in [\numagents]}\sum_{\idxalt\in\vertices_\truthful} \adjacency_{\idx\idxalt} < \sum_{\idx \in [\numagents]}\sum_{\idxalt\in\vertices_\truthful} \adjacency^\prime_{\idx\idxalt}\right) \geq 1 - \delta_1,
\end{align*}
where $\delta_1 = \exp\left(-\frac{(1-\frac{1}{\numagents})^2\log\numagents}{2(\max(\frac{1}{\numagents\problarge_\truthful},\frac{1}{\numagents\problarge_\hallucinating})-1)^2}\right)$.
\end{proposition}
\textit{Proof Outline: }
We first analyze Step 6 and bound the probability of error $\probability(\gradevector_\idxalt >\frac{\expectedgrade_\truthful+\expectedgrade_\dontknow}{2}\mid \latentstate = \truthful,\numagents-1\geq |\neighbours(\idx)|\geq\frac{4\log(2\numagents)}{|\expectedgrade_\truthful-\expectedgrade_\dontknow|^2}\}) \geq 1 - \frac{1}{\numagents}$ using Hoeffding's inequality. We then bound the probability of Step 5 succeeding (the sampled edge is from a truthful node), $\probability(\latentstate=\truthful|\gradevector_\idxalt >\frac{\expectedgrade_\truthful+\expectedgrade_\dontknow}{2},\numagents-1\geq |\neighbours(\idx)|\geq\frac{4\log(2\numagents)}{|\expectedgrade_\hallucinating-\expectedgrade_\dontknow|^2}) \geq \frac{(1-\frac{1}{\numagents})\problarge_\truthful}{\frac{1-\problarge_\truthful}{\numagents}+\problarge_\truthful} $ using Bayes' rule. A similar bound can be obtained for Step 9. A single iteration, therefore, succeeds with probability $(\frac{(1-\frac{1}{\numagents})\problarge_\truthful}{\frac{1-\problarge_\truthful}{\numagents}+\problarge_\truthful})\frac{(1-\frac{1}{\numagents})\problarge_\hallucinating}{\frac{1-\problarge_\hallucinating}{\numagents}+\problarge_\hallucinating}$. The final step is applying Hoeffding's inequality over the success of $\log \numagents$ rounds to bound the probability that more rounds succeed than not.  

The above proposition shows that the number of edges from truthful nodes increases in a single algorithm run with probability $1-\delta_1$, where  $\delta_1$ is especially small when the $\problarge_\truthful$ and $\problarge_\hallucinating$ are away from $0$ and $1$. If the $\problarge_\hallucinating$ is close to $1$, then Step 9 does not run, and one can obtain a stronger bound.

\section{Numerical Results: Effect of Network Reconfiguration and Control Optimization }
In this section, we present numerical experiments that 
demonstrate how a network of LLMs reaches consensus by information diffusion and adaptive network readjustment using Algorithm~\ref{alg:trustrankbasedreadjustment}. Then we present an example cost function that can be minimized to control the reliability and efficiency of the system. Lastly, we demonstrate how our proposed cost functions can be minimized to reduce the communication overhead and improve convergence.  
\vspace{-2mm}

\subsection{Dataset and Task Description}

We consider the task of answering a question given access to a distributed dataset, where a single LLM analyzes the associated data. The private dataset of the LLM could either contain the information required to answer the question or not. 
The dataset consists of 20 scenarios, each associated with a question, its correct answer, and the document which contains the answer. Each document is split into five paragraphs. Some paragraphs contain the correct answer to the question, while others do not, and each LLM receives a paragraph randomly during initialization\footnote{Our experimental setup, along with 
 dataset, code, LLM prompts, and supplementary results are available at \text{github.com/yimingz1218/llm\_networks}.}. 
 In our experiment, we consider a network of 100 LLMs with the open-source LLM Llama 3.1 8B. The network is represented by a directed adjacency matrix that defines each LLM's inbound and outbound neighborhood connections. Each LLM is assigned a random paragraph, with 30\% of the LLMs receiving paragraphs that contain the correct answer to a corresponding question.
To demonstrate the effectiveness of dynamically readjusting the network, LLMs are prompted at each iteration to reconsider their answers by incorporating the opinions of their neighbors. Every 200 iterations, LLMs evaluate the reliability of their neighbors (without any private context, hence ensuring independent evaluations and respecting assumption B1) and update the adjacency matrix using Algorithm~\ref{alg:trustrankbasedreadjustment}.





\subsection{Benchmark against different preferential attachment networks initializations with our reconfiguration}

\begin{figure}[ht!]
    \centering
    \includegraphics[width=\linewidth]{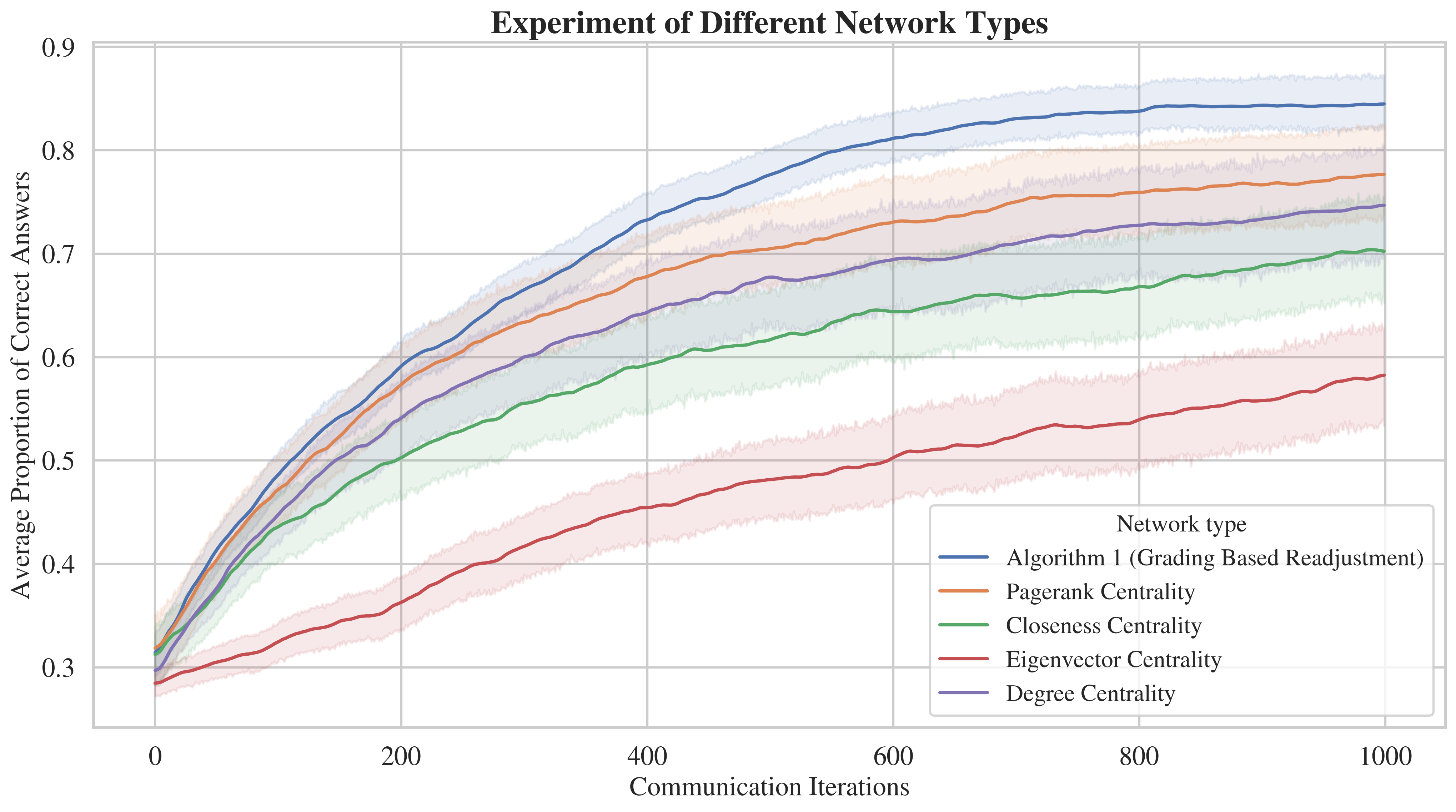}
    \caption{Algorithm~\ref{alg:trustrankbasedreadjustment} can readjust the network to ensure a faster convergence to stable equilibrium near $(1,0,0)$  compared to other static preferential attachment-based network initializations.}
    \vspace{-5mm}
    \label{fig:init}
\end{figure}

Figure~\ref{fig:init} illustrates the comparison across different preferential attachment mechanisms with our Algorithm~\ref{alg:trustrankbasedreadjustment} on the convergence. The average proportion of correct answers across 20 different questions, with five repeated experiments per question to reduce variance. The network is initialized using various centrality measures—namely, PageRank, closeness, eigenvector, and degree centrality, which do not exploit the evaluation capabilities of LLMs. The results demonstrate that our method achieves more consistent and truthful convergence across different questions. Although we are not able to achieve a locally asymptotically stable equilibrium at $(1,0,0)$ guaranteed by Theorem 1 under the conditions A1 and A2, we do achieve a stable equilibrium near it at $(0.9,0.1,0)$, and the values of $\kappa_{\truthful\hallucinating}(\control,\degreeidx,\degreeidx,0)<0.02$ for all $\degreeidx$, which motivate studying generalizations of Theorem 1.  The average grades computed were $\expectedgrade_\truthful = 6.65, \expectedgrade_\dontknow = 5.50, \expectedgrade_\hallucinating = 4.98$ validating assumption B1.

\subsection{Optimizing For Token Efficiency and Convergence}
The control $\control$ which governs the dynamics of~\eqref{eq:degreebaseddynamiceqn}, in practice, could be the token count that the LLMs used to communicate or the tokens used to process private information. Therefore, one could use the two-time scale abstraction to control the system further to achieve the inference objective while minimizing the token efficiency. If one could exactly extract the coefficient parameters of the differential equations governing the dynamics, they could use a model predictive control approach to optimize for the cost function. However, although the differential equations offer theoretical insights into the convergence, often estimating the ODE parameters for reliably such a system is computationally intensive; future work can look at sample efficient ways of learning the ODE. Instead, we show our numerical results using a simultaneous perturbation-based stochastic approximation (SPSA) approach~\cite{kushner_stochastic_2003}. In practice, the control variable $\control$ is from a finite set of integers. However, we relax the problem to a continuous domain and perform a classical SPSA-based optimization. 
We solve the following stochastic continuous optimization problem, which optimizes for the control $\control \in \controlspace$ to minimize the expected cost function $\costfunction$ where the expectation is over the network dynamics,
\begin{align*}
    \min_{\control\in\controlspace} \expectation\left\{\costfunction((\fractiontruthful^{\timeindex})_0^{\timehorizon}, (\degreedistvect^{\timeindex})_0^{\timehorizon},\control)\right\},
\end{align*}
where $\timehorizon$ is the time horizon.
For our experimental results, we consider the following cost function: 
\begin{align}\label{eq:costfunction}
   \costfunction((\fractiontruthful^{\timeindex})_0^{\timehorizon}, (\degreedistvect^{\timeindex})_0^{\timehorizon},\control) = &\constants_{\text{c}} \sum_{\timeindex=1}^\timehorizon \expectation_{\degreedistvect_\timeindex}[\costfunction_{\text{c}}(\degreeidx,\control)]- \constants_{\text{a}}(1 - \fractiontruthful_\truthful^\timehorizon)
\end{align}
where $\constants_{\text{c}}$ and $\constants_{\text{a}}$ (we take $\frac{\constants_{\text{c}}}{\constants_{\text{a}}}= 10^{-6}$) are weighing constants for the two different cost components. The first part deals with the expected communication cost using the cost function $\costfunction_{\text{c}}(\degreeidx,\control)$ for degree $\degreeidx$ and control $\control$ and the second term penalizes hallucination.

\begin{figure}[ht!]
    \centering
    \includegraphics[width=\linewidth]{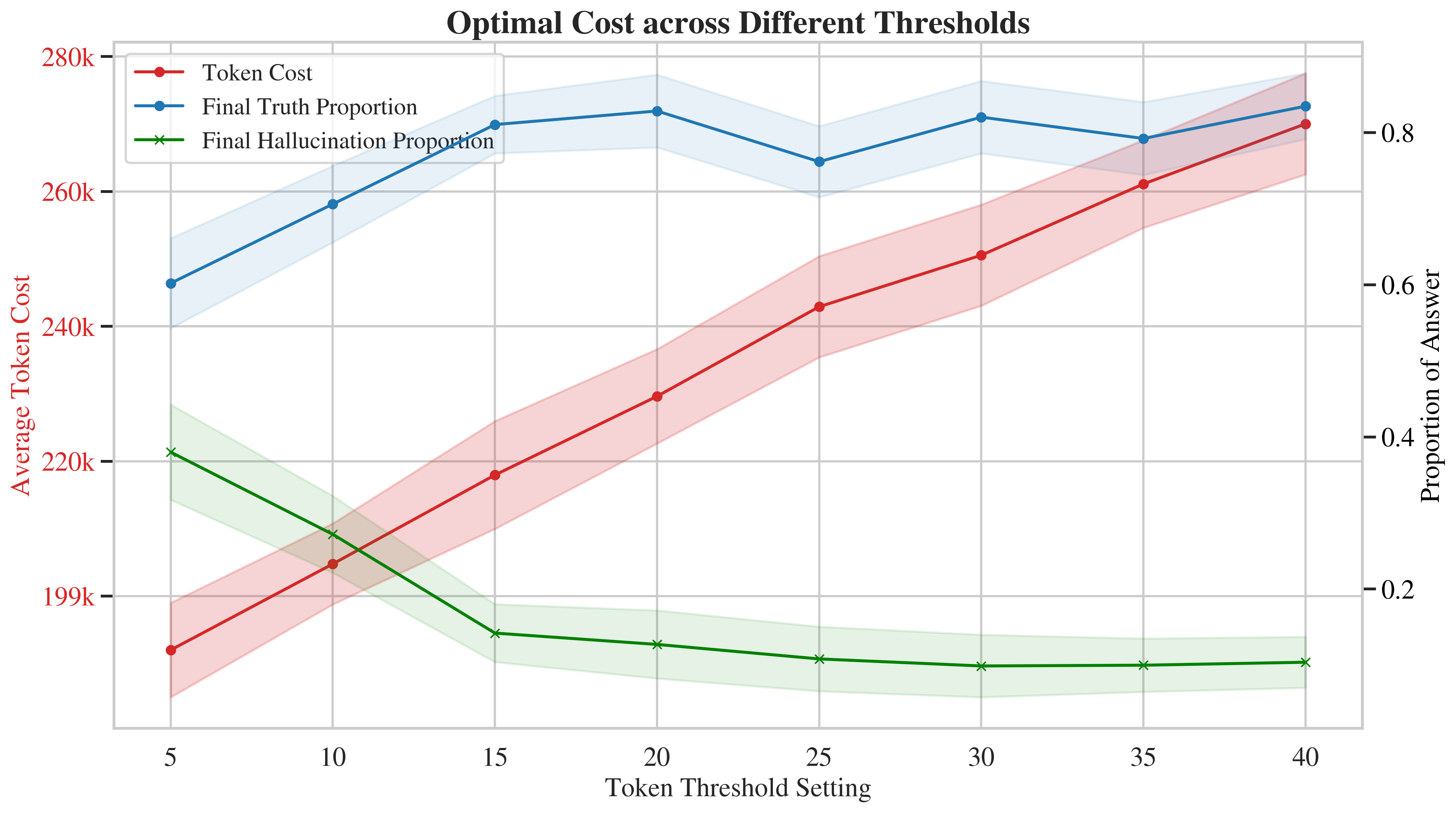}
    \caption{Final value of $\fractiontruthful_\truthful$ and token cost under varying the control $\control$ (soft token thresholds) across 20 questions (5 trials each). $\fractiontruthful_\truthful$ peaks around control $\control=20$ and $\fractiontruthful_\hallucinating$ peaks around control $\control=35$, then plateus, while token cost rises steadily. This highlights the need for optimized control to balance cost and hallucination reduction.}
    \label{fig:cost}
\end{figure}

In Figure~\ref{fig:cost}, we report the average token cost, the proportion of truthful LLMs ($\fractiontruthful_\truthful$) and the proportion of hallucinating LLMs ($\fractiontruthful_\hallucinating$) for different values of our control $\control$ across 20 distinct questions, each evaluated over five repeated trials to reduce variance. The control ($\control$) we consider is a token budget is enforced through a soft limit, wherein the prompt encourages the LLM to produce responses within a specified token range. In addition, a hard limit is imposed by constraining the output length of the \llma\ through parameter settings in code, computed as a fixed overhead plus the soft limit. Notably, increasing the token limit does not consistently enhance performance. 
This observation motivates optimizing the cost function defined in \eqref{eq:costfunction}, aiming to achieve a more effective trade-off between accuracy and token efficiency.

\begin{figure}[ht!]
    \centering
    \includegraphics[width=\linewidth]{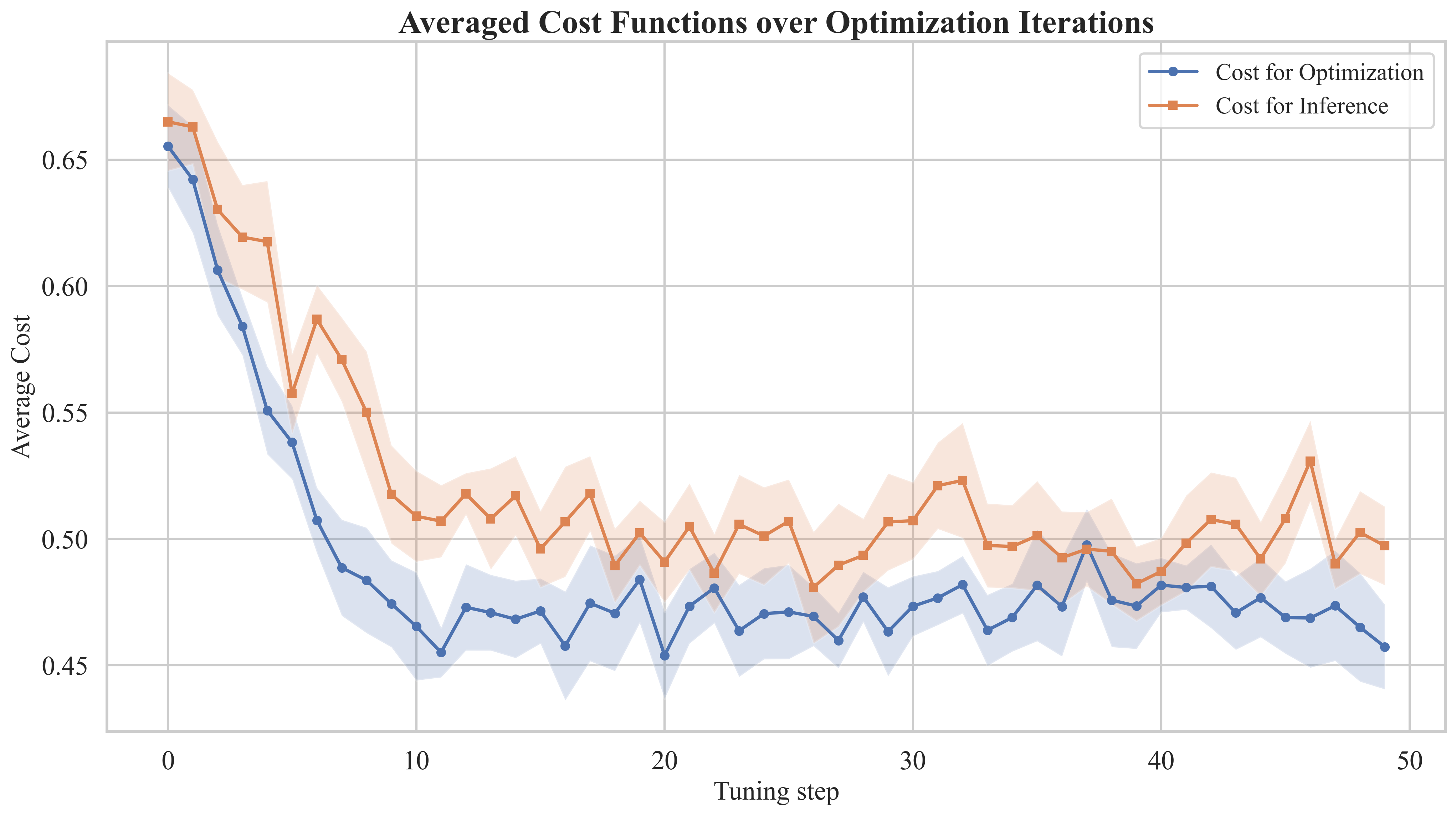}
    \caption{Gradient descent can be used to optimize the cost function associated with a network of LLMs. The blue line indicates the cost function being optimized using a set of $10$ question and answer (QA) pair set. The orange line indicates the cost function on a separate $10$ QA pair set.  The cost decreases and stabilizes, indicating effective convergence. This confirms that the optimized $\control$ successfully optimizes the trade-off defined in~\eqref{eq:costfunction}.}
    \label{fig:cost_des}
\end{figure}

To demonstrate the effectiveness of our control strategy in improving token efficiency and convergence, we adjust the token limit every 1000 iterations with a step size of 2. The direction of adjustment is determined by recalculating the average cost function over a set of 5 questions, guiding whether to increase or decrease the token limit in the next step.
As shown in Figure~\ref{fig:cost_des}, we use SPSA to optimize the cost function defined in~\eqref{eq:costfunction}. Every 1000 iterations, the token limit assigned to the LLMs is updated based on the finite gradient approximation. We start with an initial control of $\control_0=45$ and converge to a control of $ \control_{50} = 18$. The plot illustrates that the overall cost decreases over the optimization steps and converges both on the dataset the optimization is performed using and a separate inference dataset.

\section{Conclusion}
This paper proposed a mean-field approximation for information diffusion in a network of LLMs performing distributed inference.  A two-time scale abstraction for the system, which can be studied using singular perturbation methods, is provided. Further, we propose a preferential attachment protocol that uses a distributed evaluation framework to readjust the network on a slower time scale. Finally, we numerically study the efficacy of our proposed protocol and show how we can control the parameters of the interaction of individual LLMs. Future work could extend the theoretical results by studying more general equilibria beyond the simple case where all LLMs are truthful. 
\bibliographystyle{abbrv}
\bibliography{refs}

\begin{thebibliography}{10}

\bibitem{bhargava2024whatsmagicwordcontrol}
A.~Bhargava, C.~Witkowski, S.-Z. Looi, and M.~Thomson.
\newblock What's the magic word? a control theory of llm prompting, 2024.

\bibitem{s16040481}
Z.~Duan, M.~Yan, Z.~Cai, X.~Wang, M.~Han, and Y.~Li.
\newblock Truthful incentive mechanisms for social cost minimization in mobile crowdsourcing systems.
\newblock {\em Sensors}, 16(4), 2016.

\bibitem{gupta-etal-2024-systematic}
L.~Gupta, S.~Sharma, and Y.~Zhao.
\newblock Systematic evaluation of long-context {LLM}s on financial concepts.
\newblock In F.~Dernoncourt, D.~Preo{\c{t}}iuc-Pietro, and A.~Shimorina, editors, {\em Proceedings of the 2024 Conference on Empirical Methods in Natural Language Processing: Industry Track}, pages 1163--1175, Miami, Florida, US, Nov. 2024. Association for Computational Linguistics.

\bibitem{accessieee}
A.~Jain and V.~Krishnamurthy.
\newblock Interacting large language model agents {Bayesian} social learning based interpretable models.
\newblock {\em IEEE Access}, 13:25465--25504, 2025.

\bibitem{joseph2021controllability}
G.~Joseph, B.~Nettasinghe, V.~Krishnamurthy, and P.~K. Varshney.
\newblock Controllability of network opinion in erdos-renyi graphs using sparse control inputs.
\newblock {\em SIAM journal on control and optimization}, 59(3):2321--2345, 2021.

\bibitem{khalil2002nonlinear}
H.~K. Khalil and J.~W. Grizzle.
\newblock {\em Nonlinear systems}, volume~3.
\newblock Prentice hall Upper Saddle River, NJ, 2002.

\bibitem{krishnamurthy_partially_2016}
V.~Krishnamurthy.
\newblock {\em Partially {Observed} {Markov} {Decision} {Processes}}.
\newblock Cambridge University Press, Mar. 2016.

\bibitem{Krishnamurthy2019}
V.~Krishnamurthy and B.~Nettasinghe.
\newblock {\em Information Diffusion in Social Networks: Friendship Paradox Based Models and Statistical Inference}, pages 369--406.
\newblock Springer International Publishing, Cham, 2019.

\bibitem{consensusformationtwotimescale}
V.~Krishnamurthy, K.~Topley, and G.~Yin.
\newblock Consensus formation in a two-time-scale markovian system.
\newblock {\em Multiscale Modeling \& Simulation}, 7(4):1898--1927, 2009.

\bibitem{kushner_stochastic_2003}
H.~Kushner and G.~Yin.
\newblock {\em Stochastic {Approximation} and {Recursive} {Algorithms} and {Applications}}.
\newblock Stochastic {Modelling} and {Applied} {Probability}. Springer New York, 2003.

\bibitem{li2017mechanismdesignsocialnetworks}
B.~Li, D.~Hao, D.~Zhao, and T.~Zhou.
\newblock Mechanism design in social networks, 2017.

\bibitem{longcontexteval}
N.~F. Liu, K.~Lin, J.~Hewitt, A.~Paranjape, M.~Bevilacqua, F.~Petroni, and P.~Liang.
\newblock Lost in the middle: How language models use long contexts.
\newblock {\em Transactions of the Association for Computational Linguistics}, 12:157--173, 02 2024.

\bibitem{López-Pintado2016}
D.~L{\'o}pez-Pintado.
\newblock {\em An Overview of Diffusion in Complex Networks}, pages 27--48.
\newblock Springer International Publishing, Cham, 2016.

\bibitem{lopez-pintado_diffusion_2008}
D.~López-Pintado.
\newblock Diffusion in complex social networks.
\newblock {\em Games and Economic Behavior}, 62(2):573--590, 2008.

\bibitem{sahoo2024comprehensive}
P.~Sahoo, P.~Meharia, A.~Ghosh, S.~Saha, V.~Jain, and A.~Chadha.
\newblock A comprehensive survey of hallucination in large language, image, video and audio foundation models.
\newblock In {\em Findings of the Association for Computational Linguistics: EMNLP 2024}, pages 11709--11724, 2024.

\bibitem{seabra2024dynamicmultiagentorchestrationretrieval}
A.~Seabra, C.~Cavalcante, J.~Nepomuceno, L.~Lago, N.~Ruberg, and S.~Lifschitz.
\newblock Dynamic multi-agent orchestration and retrieval for multi-source question-answer systems using large language models, 2024.

\bibitem{song-etal-2024-securesql}
Y.~Song, R.~Liu, S.~Chen, Q.~Ren, Y.~Zhang, and Y.~Yu.
\newblock {S}ecure{SQL}: Evaluating data leakage of large language models as natural language interfaces to databases.
\newblock In Y.~Al-Onaizan, M.~Bansal, and Y.-N. Chen, editors, {\em Findings of the Association for Computational Linguistics: EMNLP 2024}, pages 5975--5990, Miami, Florida, USA, Nov. 2024. Association for Computational Linguistics.

\bibitem{subramanian2025smalllanguagemodelsslms}
S.~Subramanian, V.~Elango, and M.~Gungor.
\newblock Small language models (slms) can still pack a punch: A survey, 2025.

\bibitem{autotqa}
J.-P. Zhu, P.~Cai, K.~Xu, L.~Li, Y.~Sun, S.~Zhou, H.~Su, L.~Tang, and Q.~Liu.
\newblock Autotqa: Towards autonomous tabular question answering through multi-agent large language models.
\newblock {\em Proc. VLDB Endow.}, 17(12):3920–3933, Aug. 2024.

\end{thebibliography}
\appendix

\subsection{Proof of Theorem 1}
\begin{proof} 
To establish that the equilibrium \(\fractiontruthful^\degreeidx_* = [1, 0, 0]^T\) for all \(\degreeidx\) is locally asymptotically stable, we employ the Lyapunov stability theorem. This requires constructing a Lyapunov function \(\lyuponov(\fractiontruthful)\) that is positive definite near the equilibrium and demonstrating that its time derivative \(\frac{d}{d\timeindexalt} \fractiontruthful(\fractiontruthful)\) is negative definite in a neighborhood of \(\fractiontruthful\) for all $\degreeidx$.

\textbf{Step 1: Define the Lyapunov Function}
Consider the candidate Lyapunov function for each degree \(\degreeidx\):
\begin{align*}
    \lyuponov^\degreeidx(\fractiontruthful^\degreeidx) = \fractiontruthful_\hallucinating^\degreeidx + \fractiontruthful_\dontknow^\degreeidx 
\end{align*}
Since \(\fractiontruthful_\hallucinating^\degreeidx \geq 0\) and \(\fractiontruthful_\dontknow^\degreeidx  \geq 0\), we have \(\lyuponov^\degreeidx(\fractiontruthful)\geq 0\). Moreover, \(\lyuponov^\degreeidx(\fractiontruthful) = 0\) if and only if \(\fractiontruthful_\hallucinating^\degreeidx\) and \(\fractiontruthful_\dontknow^\degreeidx  = 0\), which, given the constraint \(\sum \fractiontruthful_\latentstate^\degreeidx = 1\), implies \(\fractiontruthful_\truthful^\degreeidx = 1\). Thus, \(\lyuponov^\degreeidx(\fractiontruthful^\degreeidx)  = 0\) at the equilibrium \(\fractiontruthful\degreeidx^*\).
For the entire system, define:
\[
\lyuponov(\fractiontruthful)  = \sum_\degreeidx \degreedistvect_\degreeidx \lyuponov^\degreeidx(\fractiontruthful^\degreeidx).
\]
Clearly, \(\lyuponov(\fractiontruthful) \geq 0\), and \(\lyuponov(\fractiontruthful) = 0\) if and only if \(\fractiontruthful_\hallucinating^\degreeidx = 0\) and \(\fractiontruthful_\dontknow^\degreeidx = 0\) for all \(\degreeidx\), i.e., \(\mathbf{\rho} = \mathbf{\rho}_*\). Hence, \(\lyuponov(\fractiontruthful)\) is positive definite around the equilibrium.

\textbf{Step 2: Compute the Time Derivative}
Next, compute the time derivative of the Lyapunov function:
\begin{align*}
\frac{d}{dt} \lyuponov(\fractiontruthful) =\sum_\degreeidx \degreedistvect_\degreeidx \left( \frac{d}{dt} \fractiontruthful_\hallucinating^\degreeidx + \frac{d}{dt} \fractiontruthful_\dontknow^\degreeidx \right).
\end{align*}
The specific dynamics are given as:
\begin{align*}
\frac{d}{dt} \fractiontruthful_\hallucinating^\degreeidx &= \avgprobabilitytransition{\truthful}{\hallucinating}^\degreeidx \fractiontruthful_\truthful^\degreeidx +\avgprobabilitytransition{\dontknow}{\hallucinating}^\degreeidx \fractiontruthful_\dontknow^\degreeidx - (\avgprobabilitytransition{\hallucinating}{\truthful}^\degreeidx  + \avgprobabilitytransition{\hallucinating}{\dontknow}^\degreeidx)\fractiontruthful_\hallucinating^\degreeidx,\\
\frac{d}{dt} \fractiontruthful_\dontknow^\degreeidx &= \avgprobabilitytransition{\truthful}{\dontknow}^\degreeidx \fractiontruthful_\truthful^\degreeidx +\avgprobabilitytransition{\hallucinating}{\dontknow}^\degreeidx \fractiontruthful_\hallucinating^\degreeidx - (\avgprobabilitytransition{\dontknow}{\truthful}^\degreeidx  + \avgprobabilitytransition{\dontknow}{\hallucinating}^\degreeidx)\fractiontruthful_\dontknow^\degreeidx.
\end{align*}
Summing these and plugging $\fractiontruthful_\truthful^\degreeidx = 1 - \fractiontruthful_\hallucinating^\degreeidx- \fractiontruthful_\dontknow^\degreeidx$:
\begin{align*}
\frac{d}{dt} (\fractiontruthful_\hallucinating^\degreeidx + \fractiontruthful_\dontknow^\degreeidx) &= \avgprobabilitytransition{\truthful}{\hallucinating}^\degreeidx (1 - \fractiontruthful_\hallucinating^\degreeidx- \fractiontruthful_\dontknow^\degreeidx) +\avgprobabilitytransition{\dontknow}{\hallucinating}^\degreeidx \fractiontruthful_\dontknow^\degreeidx \\&- (\avgprobabilitytransition{\hallucinating}{\truthful}^\degreeidx  + \avgprobabilitytransition{\hallucinating}{\dontknow}^\degreeidx)\fractiontruthful_\hallucinating^\degreeidx+ \avgprobabilitytransition{\truthful}{\dontknow}^\degreeidx (1 - \fractiontruthful_\hallucinating^\degreeidx- \fractiontruthful_\dontknow^\degreeidx) \\&+\avgprobabilitytransition{\hallucinating}{\dontknow}^\degreeidx \fractiontruthful_\hallucinating^\degreeidx - (\avgprobabilitytransition{\dontknow}{\truthful}^\degreeidx  + \avgprobabilitytransition{\dontknow}{\hallucinating}^\degreeidx)\fractiontruthful_\dontknow^\degreeidx
\end{align*}
And we can write the change in the Lyuponov function as, 
\begin{align*}
\frac{d}{dt} \lyuponov(\fractiontruthful) &= \sum_\degreeidx \degreedistvect_\degreeidx \left[ (\avgprobabilitytransition{\truthful}{\hallucinating}^\degreeidx +\avgprobabilitytransition{\truthful}{\dontknow}^\degreeidx) (1 - \fractiontruthful_\hallucinating^\degreeidx- \fractiontruthful_\dontknow^\degreeidx)  \right.\\&\left.- \avgprobabilitytransition{\hallucinating}{\truthful}^\degreeidx \fractiontruthful_\hallucinating^\degreeidx - \avgprobabilitytransition{\dontknow}{\truthful}^\degreeidx \fractiontruthful_\dontknow^\degreeidx   \right].
\end{align*}
\textbf{Step 3: Analyze Near the Equilibrium }
Evaluate the behavior near $\rho_*$, where $\rho_\truthful^\degreeidx = 1$, $\rho_\hallucinating^\degreeidx = 0$, $\rho_\dontknow^\degreeidx = 0$, so $\theta_\truthful = 1$, $\theta_\hallucinating = 0$, $\theta_\dontknow = 0$. The transition rate is given by~\eqref{eq:G}.

\subsubsection*{At the Equilibrium}
When $\theta_\truthful = 1$, $\theta_\hallucinating = 0$, $\theta_\dontknow = 0$, only the term with $i = \degreeidx$, $j = 0$ contributes (since $(\theta_\truthful)^\degreeidx = 1$, and $1 - \theta_\truthful - \theta_\hallucinating = 0$):
\begin{align*}
\avgprobabilitytransition{\latentstate_1}{\latentstate_2}^\degreeidx = \functiontransition_{\latentstate_1 \latentstate_2}(\control, \degreeidx, \degreeidx, 0).
\end{align*}
By (A1), $\functiontransition_{\truthful\hallucinating}(\control, \degreeidx, \degreeidx, 0) = 0$, $\functiontransition_{\truthful\dontknow}(\control, \degreeidx, \degreeidx, 0) = 0$, so $\avgprobabilitytransition{\truthful}{\hallucinating}^\degreeidx = 0$, $\avgprobabilitytransition{\truthful}{\dontknow}^\degreeidx = 0$.
By (A2), $\functiontransition_{\hallucinating}{\truthful}(\control, \degreeidx, \degreeidx, 0) > 0$, $\functiontransition_{\dontknow\truthful}(\control, \degreeidx, \degreeidx, 0) > 0$, so define:
\begin{align*}
\alpha_{\hallucinating\truthful}^\degreeidx = \functiontransition_{\hallucinating\truthful}(\control, \degreeidx, \degreeidx, 0) > 0, \quad \alpha_{\dontknow\truthful}^\degreeidx = \functiontransition_{\dontknow\truthful}(\control, \degreeidx, \degreeidx, 0) > 0,
\end{align*}
thus $\avgprobabilitytransition{\hallucinating}{\truthful}^\degreeidx = \alpha_{\hallucinating\truthful}^\degreeidx$, $\avgprobabilitytransition{\dontknow}{\truthful}^\degreeidx = \alpha_{\dontknow\truthful}^\degreeidx$.

\subsubsection*{Near the Equilibrium}

Consider small perturbations: $\rho_\hallucinating^\degreeidx = \epsilon_\hallucinating^\degreeidx$, $\rho_\dontknow^\degreeidx = \epsilon_\dontknow^\degreeidx$, where $\epsilon_\hallucinating^\degreeidx, \epsilon_\dontknow^\degreeidx \geq 0$ are small, so $\rho_\truthful^\degreeidx = 1 - \epsilon_\hallucinating^\degreeidx - \epsilon_\dontknow^\degreeidx$. 

Let $\frac{\epsilon}{2}= \min_l \max(\epsilon_\hallucinating, \epsilon_\dontknow)$ According to A3,
\begin{align*}
\theta_\truthful = 1 - \epsilon_\hallucinating - \epsilon_\dontknow, \quad \theta_\hallucinating = \epsilon_\hallucinating, \quad \theta_\dontknow = \epsilon_\dontknow,
\end{align*}
where $\epsilon_\hallucinating, \epsilon_\dontknow$ are small constants $<\epsilon$.
We now calculate the different transition probabilities.
\begin{align*}
\avgprobabilitytransition{\truthful}{\hallucinating}^\degreeidx = \sum_{i=0}^\degreeidx \sum_{j=0}^{\degreeidx-i} \functiontransition_{\truthful\hallucinating}(\control, \degreeidx, i, j) \binom{\degreeidx}{i, j} (1 - \epsilon_\hallucinating - \epsilon_\dontknow)^i \epsilon_\hallucinating^j \epsilon_\dontknow^{\degreeidx - i - j}.
\end{align*}
Since $\functiontransition_{\truthful\hallucinating}(\control, \degreeidx, \degreeidx, 0) = 0$, the $i = \degreeidx$, $j = 0$ term is zero. Significant terms are:
$i = \degreeidx-1$, $j = 1$:
\begin{align*}
\binom{\degreeidx}{\degreeidx-1, 0} (1 - \epsilon_\hallucinating - \epsilon_\dontknow)^{\degreeidx-1} \epsilon_\dontknow = \degreeidx (1 - \epsilon_\hallucinating - \epsilon_\dontknow)^{\degreeidx-1} \epsilon_\dontknow,
\end{align*}
$i = \degreeidx-1$, $j = 1$
\begin{align*}
\binom{\degreeidx}{\degreeidx-1, 1} (1 - \epsilon_\hallucinating - \epsilon_\dontknow)^{\degreeidx-1} \epsilon_\hallucinating = \degreeidx (1 - \epsilon_\hallucinating - \epsilon_\dontknow)^{\degreeidx-1} \epsilon_\hallucinating.
\end{align*}
Next we calculate $\avgprobabilitytransition{\truthful}{\hallucinating}^\degreeidx$, $\avgprobabilitytransition{\truthful}{\dontknow}^\degreeidx$, $\avgprobabilitytransition{\hallucinating}{\truthful}^\degreeidx$ and $\avgprobabilitytransition{\dontknow}{\truthful}^\degreeidx$: 
Define $\beta_{\truthful\hallucinating}^\degreeidx = \degreeidx \functiontransition_{\truthful\hallucinating}(\control, \degreeidx, \degreeidx-1, 0)$, $\gamma_{\truthful\hallucinating}^\degreeidx = \degreeidx \functiontransition_{\truthful\hallucinating}(\control, \degreeidx, \degreeidx-1, 1)$, $\beta_{\truthful\dontknow}^\degreeidx = \degreeidx \functiontransition_{\truthful\dontknow}(\control, \degreeidx, \degreeidx-1, 0)$, $\gamma_{\truthful\dontknow}^\degreeidx = \degreeidx \functiontransition_{\truthful\dontknow}(\control, \degreeidx, \degreeidx-1, 1)$, and let $\epsilon_\truthful = 1- \epsilon_\dontknow-\epsilon_\truthful$ so:
\begin{align*}
\avgprobabilitytransition{\truthful}{\hallucinating}^\degreeidx &= \beta_{\truthful\hallucinating}^\degreeidx (\epsilon_\truthful)^{\degreeidx-1} \epsilon_\dontknow + \gamma_{\truthful\hallucinating}^\degreeidx (\epsilon_\truthful)^{\degreeidx-1} \epsilon_\hallucinating + R_{\truthful\hallucinating}^\degreeidx,\\
\avgprobabilitytransition{\truthful}{\dontknow}^\degreeidx &= \beta_{\truthful\dontknow}^\degreeidx (\epsilon_\truthful)^{\degreeidx-1} \epsilon_\dontknow + \gamma_{\truthful\dontknow}^\degreeidx (\epsilon_\truthful)^{\degreeidx-1} \epsilon_\hallucinating + R_{\truthful\dontknow}^\degreeidx,
\end{align*}
where $R_{\truthful\hallucinating}^\degreeidx$ includes higher-order terms (e.g., $i \leq \degreeidx-2$) bounded by $C_{\truthful\hallucinating}^\degreeidx (\epsilon_\hallucinating + \epsilon_\dontknow)^2$ for some constant $C_{\truthful\hallucinating}^\degreeidx$ and $R_{\truthful\dontknow}^\degreeidx \leq C_{\truthful\dontknow}^\degreeidx (\epsilon_\hallucinating + \epsilon_\dontknow)^2$.
\begin{align*}
\avgprobabilitytransition{\hallucinating}{\truthful}^\degreeidx &= \alpha_{\hallucinating\truthful}^\degreeidx (\epsilon_\truthful)^\degreeidx + \degreeidx \functiontransition_{\hallucinating\truthful}(\control, \degreeidx, \degreeidx-1, 0) (\epsilon_\truthful)^{\degreeidx-1} \epsilon_\dontknow \\&+ \degreeidx \functiontransition_{\hallucinating\truthful}(\control, \degreeidx, \degreeidx-1, 1) (\epsilon_\truthful)^{\degreeidx-1} \epsilon_\hallucinating + R_{\hallucinating\truthful}^\degreeidx,\\
\avgprobabilitytransition{\dontknow}{\truthful}^\degreeidx &= \alpha_{\dontknow\truthful}^\degreeidx (\epsilon_\truthful)^\degreeidx + \degreeidx \functiontransition_{\dontknow\truthful}(\control, \degreeidx, \degreeidx-1, 0) (\epsilon_\truthful)^{\degreeidx-1} \epsilon_\dontknow \\&+ \degreeidx \functiontransition_{\dontknow\truthful}(\control, \degreeidx, \degreeidx-1, 1) (\epsilon_\truthful)^{\degreeidx-1} \epsilon_\hallucinating + R_{\dontknow\truthful}^\degreeidx,
\end{align*}
where $R_{\hallucinating\truthful}^\degreeidx \leq C_{\hallucinating\truthful}^\degreeidx (\epsilon_\hallucinating + \epsilon_\dontknow)^2$ and $R_{\dontknow\truthful}^\degreeidx \leq C_{\dontknow\truthful}^\degreeidx (\epsilon_\hallucinating + \epsilon_\dontknow)^2$.
All these terms are needed in our computation of the next step.  
\subsubsection*{Derivative Near the Equilibrium} when $\rho_\truthful^\degreeidx  = 1 - \epsilon_\hallucinating^\degreeidx - \epsilon_\dontknow^\degreeidx$, we need to compute the following terms,
\begin{align*}
&(\avgprobabilitytransition{\truthful}{\hallucinating}^\degreeidx + \avgprobabilitytransition{\truthful}{\dontknow}^\degreeidx) (1 - \epsilon_\hallucinating^\degreeidx - \epsilon_\dontknow^\degreeidx) = \left[ (\beta_{\truthful\hallucinating}^\degreeidx + \beta_{\truthful\dontknow}^\degreeidx) \epsilon_\dontknow \right.\\&\left.+ (\gamma_{\truthful\hallucinating}^\degreeidx + \gamma_{\truthful\dontknow}^\degreeidx) \epsilon_\hallucinating + R_{\truthful\hallucinating}^\degreeidx + R_{\truthful\dontknow}^\degreeidx \right] (1 - \epsilon_\hallucinating^\degreeidx - \epsilon_\dontknow^\degreeidx) (\epsilon_\truthful)^{\degreeidx-1},\\
\end{align*}
\begin{align*}
&-\avgprobabilitytransition{\hallucinating}{\truthful}^\degreeidx \rho_\hallucinating^\degreeidx  = -\alpha_{\hallucinating\truthful}^\degreeidx (\epsilon_\truthful)^\degreeidx \epsilon_\hallucinating^\degreeidx \\&- \degreeidx \functiontransition_{\hallucinating\truthful}(\control, \degreeidx, \degreeidx-1, 0) (\epsilon_\truthful)^{\degreeidx-1} \epsilon_\dontknow \epsilon_\hallucinating^\degreeidx \\&- \degreeidx \functiontransition_{\hallucinating}{\truthful}(\control, \degreeidx, \degreeidx-1, 1) (\epsilon_\truthful)^{\degreeidx-1} \epsilon_\hallucinating \epsilon_\hallucinating^\degreeidx - R_{\hallucinating\truthful}^\degreeidx \epsilon_\hallucinating^\degreeidx,\\
-&\avgprobabilitytransition{\dontknow}{\truthful}^\degreeidx \rho_\dontknow^\degreeidx = -\alpha_{\dontknow\truthful}^\degreeidx (\epsilon_\truthful)^\degreeidx \epsilon_\dontknow^\degreeidx \\&- \degreeidx \functiontransition_{\dontknow}{\truthful}(\control, \degreeidx, \degreeidx-1, 0) (\epsilon_\truthful)^{\degreeidx-1} \epsilon_\dontknow \epsilon_\dontknow^\degreeidx \\&- \degreeidx \functiontransition_{\dontknow}{\truthful}(\control, \degreeidx, \degreeidx-1, 1) (\epsilon_\truthful)^{\degreeidx-1} \epsilon_\hallucinating \epsilon_\dontknow^\degreeidx - R_{\dontknow\truthful}^\degreeidx \epsilon_\dontknow^\degreeidx.
\end{align*}
Assume $\avgprobabilitytransition{\dontknow}{\hallucinating}^\degreeidx, \avgprobabilitytransition{\hallucinating}{\dontknow}^\degreeidx$ are bounded by $M$, $|\avgprobabilitytransition{\dontknow}{\hallucinating}^\degreeidx| \leq M$, $|\avgprobabilitytransition{\hallucinating}{\dontknow}^\degreeidx| \leq M$, so:
$|\avgprobabilitytransition{\dontknow}{\hallucinating}^\degreeidx \rho_\dontknow^\degreeidx - \avgprobabilitytransition{\hallucinating}{\dontknow}^\degreeidx \rho_\hallucinating^\degreeidx| \leq M (\epsilon_\dontknow^\degreeidx + \epsilon_\hallucinating^\degreeidx).
$
Combining and plugging $\epsilon_\truthful$ we obtain, 
\begin{align*}
\frac{d}{dt} \lyuponov(\fractiontruthful) = \sum_\degreeidx \mathbf{q}_\degreeidx&\left[ -\alpha_{\hallucinating\truthful}^\degreeidx (1 - \epsilon_\hallucinating - \epsilon_\dontknow)^\degreeidx \epsilon_\hallucinating^\degreeidx\right.\\& \left. - \alpha_{\dontknow\truthful}^\degreeidx (1 - \epsilon_\hallucinating - \epsilon_\dontknow)^\degreeidx \epsilon_\dontknow^\degreeidx + P_\degreeidx \right],
\end{align*}
where $P_\degreeidx$ includes positive and negative terms:
\begin{align*}
P_\degreeidx &= (\avgprobabilitytransition{\truthful}{\hallucinating}^\degreeidx + \avgprobabilitytransition{\truthful}{\dontknow}^\degreeidx) (1 - \epsilon_\hallucinating^\degreeidx - \epsilon_\dontknow^\degreeidx) \\&- \degreeidx \functiontransition_{\hallucinating}{\truthful}(\control, \degreeidx, \degreeidx-1, 0) (1 - \epsilon_\hallucinating - \epsilon_\dontknow)^{\degreeidx-1} \epsilon_\dontknow \epsilon_\hallucinating^\degreeidx \\&- \degreeidx \functiontransition_{\hallucinating}{\truthful}(\control, \degreeidx, \degreeidx-1, 1) (1 - \epsilon_\hallucinating - \epsilon_\dontknow)^{\degreeidx-1} \epsilon_\hallucinating \epsilon_\hallucinating^\degreeidx \\&- R_{\hallucinating\truthful}^\degreeidx \epsilon_\hallucinating^\degreeidx - \degreeidx \functiontransition_{\dontknow}{\truthful}(\control, \degreeidx, \degreeidx-1, 0) (1 - \epsilon_\hallucinating - \epsilon_\dontknow)^{\degreeidx-1} \epsilon_\dontknow \epsilon_\dontknow^\degreeidx \\& - \degreeidx \functiontransition_{\dontknow}{\truthful}(\control, \degreeidx, \degreeidx-1, 1) (1 - \epsilon_\hallucinating - \epsilon_\dontknow)^{\degreeidx-1} \epsilon_\hallucinating \epsilon_\dontknow^\degreeidx \\&- R_{\dontknow\truthful}^\degreeidx \epsilon_\dontknow^\degreeidx + (\avgprobabilitytransition{\dontknow}{\hallucinating}^\degreeidx \rho_\dontknow^\degreeidx - \avgprobabilitytransition{\hallucinating}{\dontknow}^\degreeidx \rho_\hallucinating^\degreeidx).
\end{align*}

Since $\epsilon_\hallucinating, \epsilon_\dontknow, \epsilon_\hallucinating^\degreeidx, \epsilon_\dontknow^\degreeidx$ are small, and from the previous step $\avgprobabilitytransition{\latentstate_1}{\latentstate_2}^\degreeidx$ for $\latentstate_2\neq \truthful$ are $\orderof(\epsilon)$, $P_\degreeidx$ is of order $\orderof(\epsilon^2)$ (products of small terms), while the leading terms $-\alpha_{\hallucinating\truthful}^\degreeidx (1 - \epsilon_\hallucinating - \epsilon_\dontknow)^\degreeidx \epsilon_\hallucinating^\degreeidx - \alpha_{\dontknow\truthful}^\degreeidx (1 - \epsilon_\hallucinating - \epsilon_\dontknow)^\degreeidx \epsilon_\dontknow^\degreeidx$ are of order $\orderof(\epsilon)$. For small perturbations ($\epsilon_\hallucinating + \epsilon_\dontknow < 1$), $(1 - \epsilon_\hallucinating - \epsilon_\dontknow)^\degreeidx > 0$, and since $\alpha_{\hallucinating\truthful}^\degreeidx > 0$, $\alpha_{\dontknow\truthful}^\degreeidx > 0$, the negative terms dominate when $\epsilon_\hallucinating^\degreeidx > 0$ or $\epsilon_\dontknow^\degreeidx > 0$.
Define $m = \min_\degreeidx \{\alpha_{\hallucinating\truthful}^\degreeidx, \alpha_{\dontknow\truthful}^\degreeidx\}$, $M_P = \max_\degreeidx \{|P_\degreeidx| / (\epsilon_\hallucinating^\degreeidx + \epsilon_\dontknow^\degreeidx)\}$. Then:
\begin{align*}
\frac{d}{dt} \lyuponov(\fractiontruthful) \leq \sum_\degreeidx \mathbf{q}_\degreeidx \left[ -m (\epsilon_\truthful)^\degreeidx (\epsilon_\hallucinating^\degreeidx + \epsilon_\dontknow^\degreeidx) + M_P (\epsilon_\hallucinating^\degreeidx + \epsilon_\dontknow^\degreeidx) \right].
\end{align*}
For $\epsilon_\hallucinating + \epsilon_\dontknow < \delta < 1$, $\epsilon_\truthful^\degreeidx = (1 - \epsilon_\hallucinating - \epsilon_\dontknow)^\degreeidx \geq (1 - \delta)^\degreeidx$, and $M_P \to 0$ as $\epsilon_\hallucinating, \epsilon_\dontknow \to 0$. Choose $\delta$ small enough that $m (1 - \delta)^\degreeidx > M_P$, ensuring for a small region around $\fractiontruthful_*$:
\begin{align*}
\frac{d}{dt} \lyuponov(\fractiontruthful) < 0 \text{ for } \fractiontruthful \neq \fractiontruthful_*.
\end{align*}
The Lyapunov function \(\lyuponov(\fractiontruthful) = \sum_\degreeidx \degreedistvect_\degreeidx (\fractiontruthful_\hallucinating^\degreeidx + \fractiontruthful_\dontknow^\degreeidx)\) is positive near \(\fractiontruthful_*^\degreeidx\), and its derivative \(\frac{d}{d\timeindexalt} \lyuponov(\fractiontruthful) < 0\) for \(\fractiontruthful \neq \fractiontruthful_*\) in a neighborhood of the equilibrium, under the given conditions. By the Lyapunov stability theorem, the equilibrium \(\fractiontruthful_* = [1, 0, 0]^\transpose \) for all \(\degreeidx\) is locally asymptotically stable.
\end{proof}
\end{document}